\documentclass[fleqn,10pt]{wlscirep}
\usepackage[T1]{fontenc}
\usepackage{bm,graphicx,color}

\newcommand{\del}[1]{}

\title{Independent engineering of individual plasmon modes in plasmonic dimers with conductive and capacitive coupling}

\author[1,2,*]{V.~K\v{r}\'apek}
\author[3]{A.~Kone\v{c}n\'a}
\author[1]{M.~Hor\'ak}
\author[1]{F.~Ligmajer}
\author[4]{M.~St\"{o}ger-Pollach}
\author[1]{M.~Hrto\v{n}}
\author[1]{J.~Babock\'y}
\author[1,2]{T.~\v{S}ikola}

\affil[1]{Central European Institute of Technology, Brno University of Technology, Purky\v{n}ova 123, 612 00 Brno, Czech Republic}
\affil[2]{Institute of Physical Engineering, Brno University of Technology, Technick\'a 2, 616 69 Brno, Czech Republic}
\affil[3]{Materials Physics Center CSIC-UPV/EHU, Paseo Manuel de Lardizabal 5, 20018 San Sebasti\'an, Spain}
\affil[4]{University Service Centre for Transmission Electron Microscopy, Vienna University of Technology, Wiedner Hauptstra{\ss}e 8--10, A-1040 Wien, Austria}
\affil[*]{vlastimil.krapek@ceitec.vutbr.cz}

\del{
\author{V.~K\v{r}\'apek}
\alsoaffiliation{Institute of Physical Engineering, Brno University of Technology, Technick\'a 2, 616 69 Brno, Czech Republic}
\affiliation{Central European Institute of Technology, Brno University of Technology, Purky\v{n}ova 123, 612 00 Brno, Czech Republic}
\email{vlastimil.krapek@ceitec.vutbr.cz}
\author{A.~Kone\v{c}n\'a}
\affiliation{Materials Physics Center CSIC-UPV/EHU, Paseo Manuel de Lardizabal 5, 20018 San Sebasti\'an, Spain}
\author{M.~Hor\'ak}
\author{F.~Ligmajer}
\affiliation{Central European Institute of Technology, Brno University of Technology, Purky\v{n}ova 123, 612 00 Brno, Czech Republic}
\author{M.~St\"{o}ger-Pollach}
\affiliation{University Service Centre for Transmission Electron Microscopy, Vienna University of Technology, Wiedner Hauptstra{\ss}e 8--10, A-1040 Wien, Austria}
\author{M.~Hrto\v{n}}
\author{J.~Babock\'y}
\author{T.~\v{S}ikola}
\alsoaffiliation{Institute of Physical Engineering, Brno University of Technology, Technick\'a 2, 616 69 Brno, Czech Republic}
\affiliation{Central European Institute of Technology, Brno University of Technology, Purky\v{n}ova 123, 612 00 Brno, Czech Republic}
}

\keywords{Plasmonics; localized surface plasmons; electron energy loss spectroscopy; Babinet's principle; hot spots}

\begin{abstract}

We revisit plasmonic modes in nanoparticle dimers with conductive or insulating junction resulting in conductive or capacitive coupling. In our study which combines electron energy loss spectroscopy, optical spectroscopy, and numerical simulations, we show coexistence of strongly and weakly hybridized modes. While the properties of the former ones strongly depend on the nature of the junction, the properties of the latter ones are nearly unaffected. This opens up a prospect for independent engineering of different plasmonic modes in a single plasmonic antenna. In addition, we show that Babinet's principle allows to engineer the near field of plasmonic modes independent of their energy.
Finally, we demonstrate that combined electron energy loss imaging of a plasmonic antenna and its Babinet-complementary counterpart allows to reconstruct the distribution of both electric and magnetic near fields of localised plasmonic resonances supported by the antenna as well as charge and current antinodes of related charge oscillations.

\end{abstract}

\begin{document}

\flushbottom
\maketitle
\thispagestyle{empty}

\section*{Introduction}

Plasmonic antennas are metallic particles widely studied for their ability to control, enhance, and concentrate electromagnetic field~\cite{Novotny2011}. Strikingly, the field in the vicinity of the plasmonic antennas, the so-called near field, can be focused to deeply subwavelength region. At the same time, the field is strongly enhanced (in comparison to the driving field, which can be e.g. a plane wave).  Focusing of the field stems from the excitation of localized surface plasmons (LSP) -- quantized oscillations of the free electron gas in the metal coupled to the evanescent electromagnetic wave propagating along the boundary of the metal. Plasmonic antennas are promising for applications such as energy harvesting, non-linear optics~\cite{Klein502,Feth:08}, sensing~\cite{doi:10.1063/1.4918531}, enhanced emission~\cite{Kinkhabwala2009}, or as building blocks of metasurfaces~\cite{Yu2014}.

Properties of LSP can be engineered by adjusting the shape, size, and composition of the plasmonic antennas, by changing their dielectric environment, or by composing single antennas into more complex structures -- dimers and multimers. LSP modes in the dimers can be described in terms of hybridization~\cite{doi:10.1021/nl049681c,doi:10.1021/nl101335z}: the modes of individual antennas combine into bonding and antibonding hybridized modes. For large enough separation between the components of the dimer antenna, the coupling between the modes can be described as a dipole-dipole coupling~\cite{doi:10.1021/nl049681c}. For closely spaced components, more complex hybridization models exist involving multipolar terms~\cite{doi:10.1021/jp0613485} or near-field coupling~\cite{doi:10.1021/nl101335z}. Instead of this capacitive coupling, the components can be coupled conductively by a metallic junction. The effects of the transformation between the capacitive and conductive coupling have been thoroughly studied~\cite{doi:10.1021/nl104410t,doi:10.1021/nl3001309}, indicating pronounced qualitative and quantitative changes in the properties of LSP modes, going far beyond the simple hybridization concept in the case of a conductive coupling.

Babinet's principle represents another concept for engineering LSP modes. It relates optical response of thin solid and hollow antennas: metallic particles and apertures in the metallic films of the same size and shape. Both types of antennas provide complementary response with identical energies of LSP modes but with interchanged electric and magnetic field~\cite{born1999principles,PhysRevLett.93.197401,PhysRevB.76.033407,doi:10.1021/nl402269h}. Applicability of Babinet's principle in plasmonics over a broad spectral range from THz to visible has been experimentally demonstrated~\cite{Bitzer:11,Rossouw:12,Mizobata:17,vonCube:11} with quantitative deviations being observed, in particular at optical frequencies~\cite{Mizobata:17,Horak2018SciRep}. Babinet's principle has been used to design plasmonic structures with advanced functionalities including metasurfaces and metamaterials~\cite{PhysRevLett.93.197401,Chen:07} or antennas with electromagnetic hot spots~\cite{Hrton2019arxiv}. It is also used as a tool for indirect imaging of magnetic field. Here, the magnetic field distribution formed by a planar object of interest is first related through Babinet's principle to the electric field distribution for a complementary structure (interchanging conducting and insulating parts), which is considerably easier to detect. The electric field distribution is then retrieved by one of available methods such as the electron energy loss spectroscopy (EELS)~\cite{doi:10.1021/acsphotonics.6b00857} or near-field microscopy~\cite{Bitzer:11}. Babinet's reciprocity between electric and magnetic field has been also used to interchange electric and magnetic hot spots~\cite{Hrton2019arxiv}.

Several interesting phenomena and applications of plasmonic antennas involve more than one LSP mode. For example, photoluminescence enhancement requires enhancement at the energies of the excitation and emission and also suppressed coupling of the emitter to non-radiative LSP modes which quench the emission~\cite{C7CS00155J}. Directional scattering is realized through the interference of the fundamental and excited LSP modes~\cite{doi:10.1021/nn502616k}. Interference of bright and dark LSP modes results into Fano resonances~\cite{doi:10.1021/nl9001876} related to plasmon-induced transparency~\cite{PhysRevLett.101.047401} or sensing~\cite{doi:10.1021/nn103172t}. It is thus desirable to have a tool for independent engineering of individual plasmonic modes in single antenna.

Here we present such a tool. We show that by switching between the capacitive and conductive coupling in bow-tie antennas, the properties of some LSP modes are changed pronouncedly while other modes are almost unaffected. We also show that by employing Babinet's principle it is possible to modify the near field of LSP modes while preserving their energy. This demonstrates feasibility of individual engineering of LSP modes in single plasmonic antennas.

\section*{Results and discussion}






Plasmonic antennas of four different types -- bowtie, diabolo, inverted bowtie, and inverted diabolo -- have been fabricated by focused ion beam lithography electron or electron beam lithography (see Methods for details) and characterized by electron energy loss spectroscopy, cathodoluminescence, transmission spectrocopy, and electromagnetic simulations.

\begin{figure}[h!]
  \begin{center}
    \includegraphics[width=\linewidth]{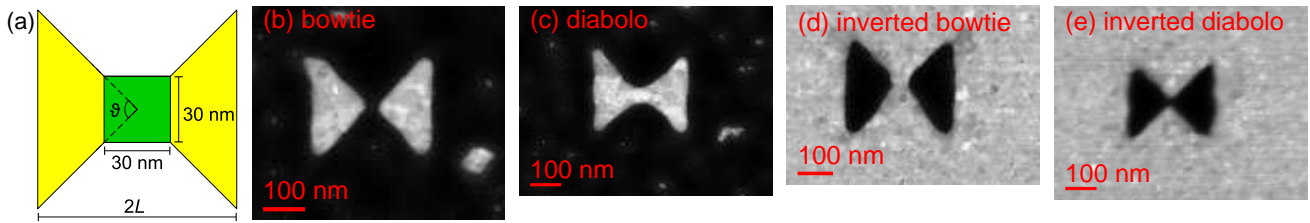}
    \caption{\label{figNO1} (a) Schematic representation of the bowtie and diabolo antennas: wings (yellow), bridge (green). $L$ denotes the length of the wing. (b--e) High-angle annular dark field transmission electron micrographs of (b) bowtie, (c) diabolo, (d) inverted bowtie, and (e) inverted diabolo antennas. Bright/dark color corresponds to gold/substrate, respectively.}
  \end{center}
\end{figure}

Figure~\ref{figNO1}(a) shows the design of the bowtie and diabolo plasmonic antennas. They consist of wings (metallic for the bowtie and diabolo) and a bridge which is insulating for a bowtie (referred to as the gap in such a case) and metallic for a diabolo. For inverted structures, metallic and insulating parts are interchanged. The dimensions of the bridge are set to 30~nm $\times$ 30~nm and the wing length $L$ is varied. The wing angle $\vartheta$ is set to $90^\circ$ which results into $1:1$ rate of metal and insulator in the active zone of the antennas and ensures maximum complementarity of direct and inverted antennas. Bowtie and inverted diabolo antennas concentrate charge and form electric hot spots (areas of high field concentration) while diabolo and inverted bowtie antennas funnel current and form magnetic hot spots~\cite{Hrton2019arxiv}. Typical fabricated antennas are shown in Fig.~\ref{figNO1}(b--e). Their clearly visible polycrystalline character shall not have profound influence on the plasmonic properties, as shown by former studies~\cite{doi:10.1021/nl3001309}.

\begin{figure}[h!]
  \begin{center}
    \includegraphics[clip,width=0.75\linewidth]{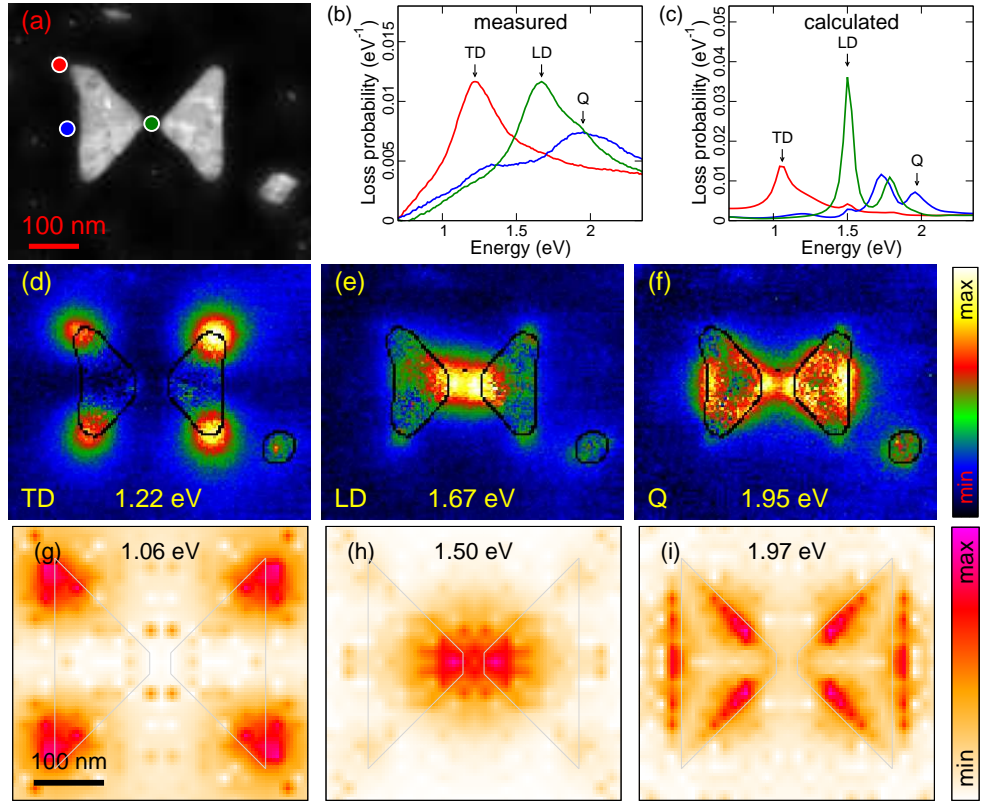}
    \caption{\label{figNO2} Bowtie antenna with the wing length $L=141$~nm. (a) HA ADF TEM image. (b,c) EEL spectra recorded at specific points indicated by color spots in panel a: (b) measured, (c) calculated. (d,e,f) Experimental energy-filtered maps of the loss intensity for (d) transverse dipole mode(TD, $E=1.23$~eV), (e) longitudinal dipole mode (LD, $E=1.67$~eV), (f) quadrupole mode (Q, $E=1.95$~eV). (g,h,i) Calculated loss intensity maps at the energy of (g) transverse dipole mode (1.06~eV), (h) longitudinal dipole mode (1.50~eV), (i) quadrupole mode (1.97~eV). Mode energies are indicated. Contours of the antennas are shown by black (experiment) or gray (calculations) lines. \del{Arrows schematically depict plasmon oscillations.}}
  \end{center}
\end{figure}

First we will discuss plasmonic properties of bowtie antennas. Typical bowtie with the wing size of $141\pm 5$~nm is shown in Fig.~\ref{figNO2}(a). EEL spectra recorded at different positions reveal three distinct peaks at the energies of 1.23~eV, 1.67~eV, and 1.95~eV, as observed in Fig.~\ref{figNO2}(b). Experimental spectra are reasonably well reproduced with calculations\del{ broadened by the instrumental spectral profile (Gaussian with the full width at half maximum of \cc{XXX})} [Fig.~\ref{figNO2}(c)]. The nature of the modes is revealed in the energy filtered maps of the loss intensity. The loss intensity is related to the plasmon electric field parallel with the trajectory of the electron beam $E_z$~\cite{RevModPhys.82.209}, which is the largest at the charge antinodes (current nodes) of plasmon oscillations. Due to large wing angle ($90^\circ$) the lowest mode at 1.23~eV is the transverse dipole mode with the charge oscillating in the direction perpendicular to the antenna's long axis. The mode is actually a doublet consisting of parallel and antiparallel alingment of the dipoles in both wings (bonding and antibonding transverse mode)~\cite{doi:10.1021/nl104410t}. However, due to large separation between the dipoles, the energy separation of both modes is small, considerably smaller than their spectral width. The next peak at 1.67~eV again comprises two modes, longitudinal dipole bonding and antibonding modes. The former features two loss intensity maxima at the bases of the wings, the latter features an additional maximum in the gap of the antenna. The third peak at 1.95~eV corresponds to the quadrupole mode. All these modes have been observed previously~\cite{doi:10.1021/nl3001309}.


\begin{figure}[h!]
  \begin{center}
    \includegraphics[clip,width=\linewidth]{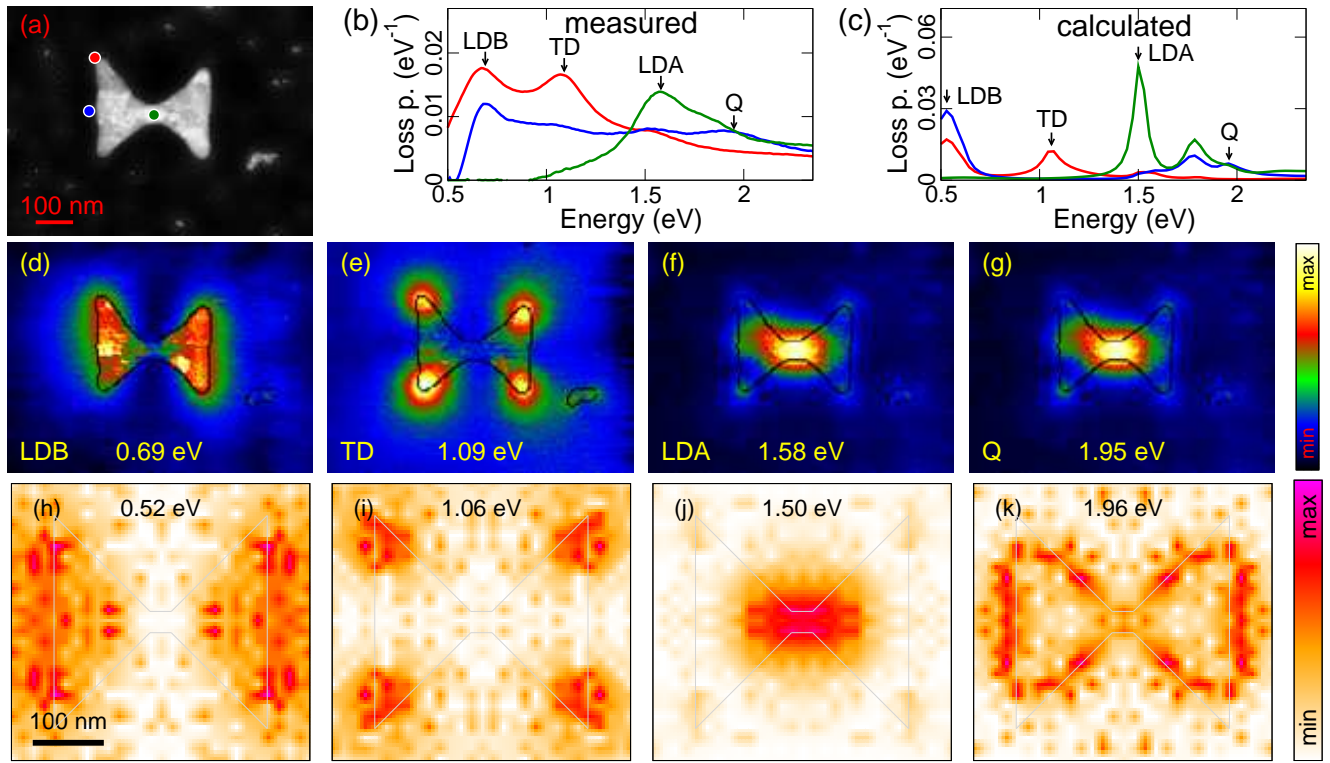}
    \caption{\label{figNO3} Diabolo antenna with the wing length $L=150$~nm. (a) HA ADF TEM image. (b,c) EEL spectra recorded at specific points indicated by color spots in panel a: measured (b), calculated (c). (d,e,f,g) Experimental energy-filtered maps of the loss intensity for (d) longitudinal bonding mode, (e) transverse mode, (f) longitudinal antibonding mode, (g) quadrupole mode. (h,i,j,k) Calculated loss intensity at the energy of (h) longitudinal bonding mode, (i) transverse mode, (j) longitudinal antibonding mode, (k) quadrupole mode. Mode energies are indicated. Contours of the antennas are shown by black (experiment) or gray (calculations) lines.}
  \end{center}
\end{figure}

We repeat similar analysis for the diabolo antenna with nearly the same wing length $L=150 \pm 5$~nm (differences arise due to limited accuracy of the fabrication). The antenna and its spectral response are shown in Fig.~\ref{figNO3}. The presence of the conductive bridge significantly influences the longitudinal bonding mode, which exhibits a huge red shift to 0.69~eV (from 1.67~eV in the bowtie) and becomes the lowest mode. The strong impact of the bridge conductivity is related to the fact the mode exhibits a current antinode in the bridge. On the contrary, the remaining modes resolved in our EELS experiment exhibit a current node within the bridge and are therefore not very sensitive to the presence of the bridge. Their energies read for the transverse dipole mode 1.09~eV (1.23~eV in the bowtie), for the longitudinal antibonding mode 1.58~eV (1.67~eV), and for the quadrupole mode 1.95~eV (1.95~eV). Weak red shifts of diabolo modes with respect to corresponding bowtie modes can be attributed to slightly larger size of the diabolo antenna involved in the comparison.

\del{  
\begin{figure}[h!]
  \begin{center}
    \includegraphics[clip,width=\linewidth]{figure3X.eps}
    \caption{\label{figNO3a}(a) Energies of plasmonic modes in bowtie (solid lines, full symbols) and diabolo (dashed lines, empty symbols) antennas as functions of the reciprocal wing length $L$: longitudinal dipolar bonding (LDB) mode (magenta), transverse dipolar (TQ) mode (orange), longitudinal dipolar antibonding (LDA) mode (maroon), quadrupolar (Q) mode (green). Energies retrieved by EELS are shown by squares, energies from optical spectroscopy are shown by circles. (b\del{,c}) Optical transmission spectra of arrays of bowtie \del{(b) and diabolo (c) }antennas for several values of the wing length. (c,d) Experimental (solid lines, full symbols) and calculated (dashed lines, empty symbols) energies of plasmonic modes for the bowtie (c) and diabolo (d) antennas as functions of the reciprocal wing length.}
  \end{center}
\end{figure}
}

\begin{figure}[h!]
  \begin{center}
    \includegraphics[clip,width=\linewidth]{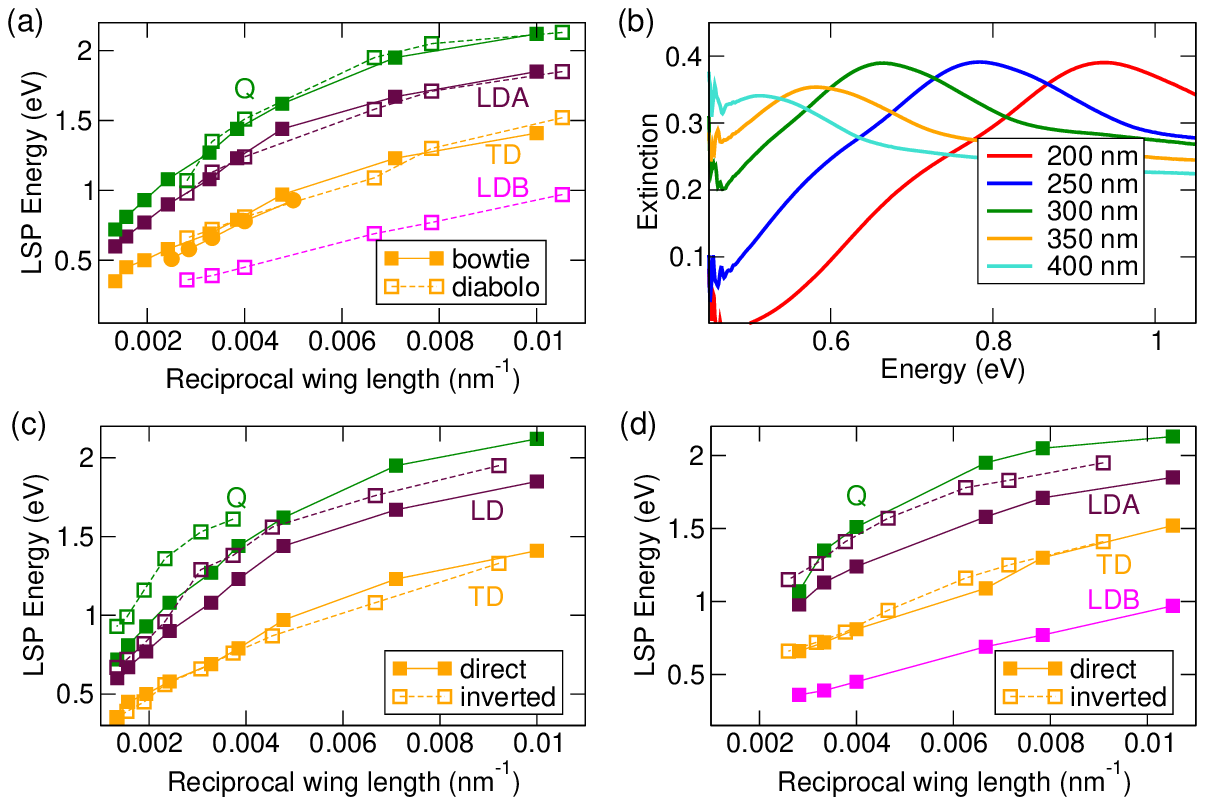}
    \caption{\label{figNO4}(a) Energies of plasmonic modes in bowtie (solid lines, full symbols) and diabolo (dashed lines, empty symbols) antennas as functions of the reciprocal wing length $L$: longitudinal dipolar bonding (LDB) mode (magenta), transverse dipolar (TQ) mode (orange), longitudinal dipolar antibonding (LDA) mode (maroon), quadrupolar (Q) mode (green). Energies retrieved by EELS are shown by squares, energies from optical spectroscopy are shown by circles. (b) Optical transmission spectra of arrays of bowtie antennas for several values of the wing length. (c) Dispersion relation for the bowtie and inverted bowtie: energies of plasmon modes as functions of reciprocal wing length. (d) Dispersion relation for the diabolo and inverted diabolo: energies of plasmon modes as functions of reciprocal wing length.}
  \end{center}
\end{figure}

Figure~\ref{figNO4}(a) compares the energies of the modes in bowtie and diabolo antennas of different sizes. Strikingly, energies of the corresponding modes in both type of antennas are nearly identical, except for the longitudinal dipolar bonding which is pronouncedly redshifted for the diabolo antenna. There is an intuitive explanation of this observation.

The modes in the dimer plasmonic antennas (such as bowtie and diabolo) can be classified as two different types. Strongly coupled modes (such as the longitudinal bonding mode) extend to the area of the junction. Depending on the nature of the junction there is either charge transfer through the conducting bridge or capacitive charge coupling (accumulation of opposite charges at opposite sides of the insulating gap). Properties of the strongly coupled modes pronouncedly depend on the properties of the junction. As an example, the longitudinal bonding mode of the bowtie antenna exhibits an electric field hotspot at the junction while the strongly red-shifted longitudinal bonding mode of the diabolo antenna exhibits a magnetic field hotspot. Weakly coupled modes (longitudinal antibonding, transverse, and quadrupolar) avoid the area of the junction, with no charge transfer or capacitive coupling, although there still might be charge accumulation (e.g. for the longitudinal antibonding mode). The junction affects the properties of the weakly coupled modes only slightly by modifying the dielectric environment and having some impact on the spatial distribution of the near field. However, charge oscillations are not directly influenced.


We note that in addition to EELS we also performed optical transmission spectroscopy for arrays of bowtie antennas [Fig.~\ref{figNO4}(b)] which exhibit only a single resolvable spectral feature corresponding to the transverse dipolar mode. Excellent agreement between the energies of the transverse dipolar modes obtained by both methods supports our interpretation of the spectral features.  

\del{Figures~\ref{figNO3a}(c) and (d) compare the experimental and calculated energies of plasmonic modes for the bowtie and diabolo antennas, respectively. The differences up to 0.1~eV for the lowest mode and up to 0.2~eV for the higher modes are within the experimental error.}

Experimental and calculated energies of plasmonic modes agree very well to each other for both the bowtie and diabolo antennas. The differences up to 0.1~eV for the lowest mode and up to 0.2~eV for the higher modes are within the experimental error. 

\del{
\begin{figure}[h!]
  \begin{center}
    \includegraphics[width=\linewidth]{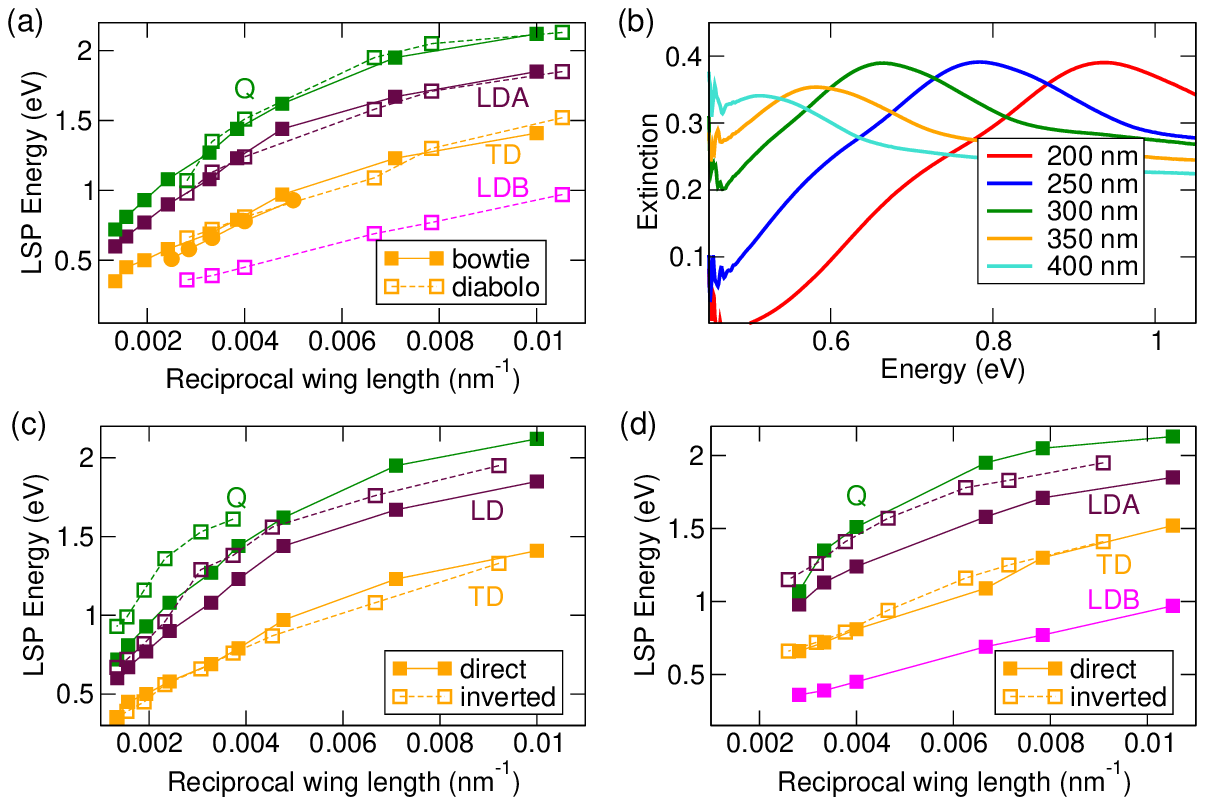}
    \caption{\label{figNO4} (a) Experimental EEL spectra for the bowtie antenna with the wing length $L=141$~nm (solid lines) and the inverted bowtie antenna with the wing length $L=150$~nm (dashed lines). (b) Experimental EEL spectra for the diabolo antenna with the wing length $L=150$~nm (solid lines) and the inverted diabolo antenna with the wing length $L=140$~nm (dashed lines). (c) Dispersion relation for the bowtie and inverted bowtie: energies of plasmon modes as functions of reciprocal wing length. (d) Dispersion relation for the diabolo and inverted diabolo: energies of plasmon modes as functions of reciprocal wing length.}
  \end{center}
\end{figure}
}

\begin{figure}[h!]
  \begin{center}
    \includegraphics[clip,width=\linewidth]{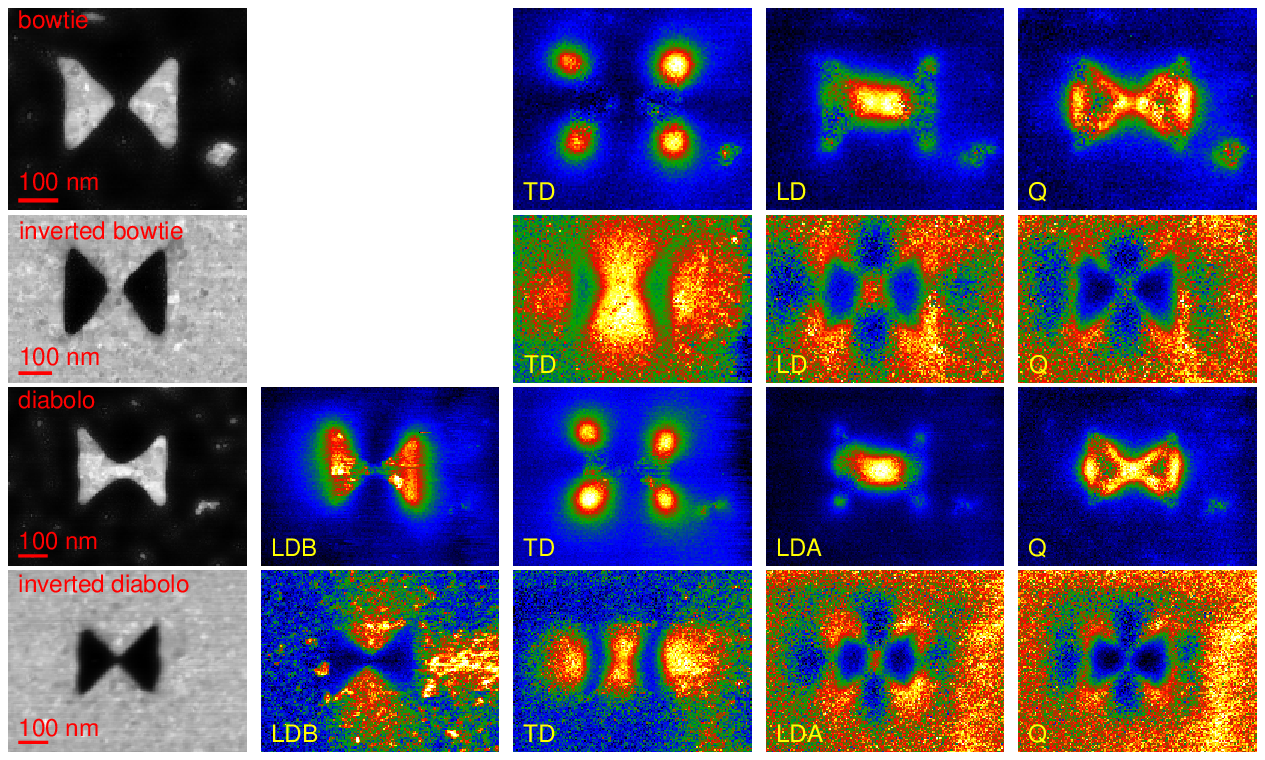}
    \caption{\label{figNO5} Transmission electron micrographs of the antennas of all four types and spatial maps of the loss function at the energy corresponding to specific LSP modes: longitudinal dipolar bonding (LDB), antibonding (LDA), and, in the case of bowtie antennas, unresolved (LD), transverse dipolar (TD), and quadrupolar (Q). The energy of the modes increases from left to right. The wing length of all antennas is comparable: 140~nm (bowtie), 150~nm (diabolo), 150~nm (inverted bowtie), 140~nm (inverted diabolo).}
  \end{center}
\end{figure}

Next, we compare direct and indirect antennas related by Babinet's principle of complementarity. \del{Typical EEL spectra plotted in Fig.~\ref{figNO4}(a,b) show that complementary antennas exhibit resonances at similar energies but with different spatial distribution of the loss function.} Inspection of the EEL spectra shows that complementary antennas exhibit resonances at similar energies but with different spatial distribution of the loss function (see Fig.~\ref{figNO5}). To compensate for the size differences we plot the peak energies of the EEL spectra as functions of the reciprocal wing length in Fig.~\ref{figNO4}(c,d) for bowtie and diabolo antennas, respectively. The interpolation between the experimentally retrieved points allows us to compare the energies of LSPR at identical sizes of complementary antennas. We clearly observe the validity of Babinet's principle: corresponding LSPR have identical energy (with respect to the experimental error, which is about 0.1~eV for the lowest modes and increases for the higher modes due to their overlapping in the spectra).

In Figure~\ref{figNO5} we compare spatial maps of the loss function for specific LSP modes of all four antenna types. The size of the antennas is identical with respect to the fabrication uncertainty and varies between 140~nm and 150~nm (wing length). We clearly observe bowtie-diabolo duality. The modes that do not involve charge transfer through the neck of the antenna (TD, LDA, Q\del{\cc{check Q}}) have identical spatial distribution of the loss function (and therefore of the near field). They also have similar mode energies. On the other hand, the mode LDB that includes charge transfer through the bridge of diabolo antennas has significantly different energy in bowtie antennas. Due to its spectral overlap with LDA it was not possible to record its spatial map experimentally. The same duality holds for the direct as well and inverted bowtie-diabolo pairs.

Babinet complementarity is also clearly visible in Fig.~\ref{figNO5}. The corresponding modes in direct and inverted antennas, despite having identical energy, strongly differ in the spatial distribution of the near field.

\del{\begin{figure}[h!]
  \begin{center}
    \includegraphics[clip,width=\linewidth]{figure6.eps}
    \caption{\label{figNO6} (a) Schematic representation of the LDB mode in the diabolo antenna. The black arrow depicts the current with the thicker middle part representing the current antinode and the thinner border parts representing the current node. Magnetic field in the plane of the antenna is schematically represented by $\odot$ (outward field) and $\otimes$ (inward field) symbols with the size corresponding to the magnitude of the field. The node of current corresponds to the accumulation of the charge represented by $+$ and $-$ symbols. (b) Experimental loss intensity distribution. (c) Calculated out-of-plane electric field. (d) Experimental loss intensity distribution for Babinet-complementary antenna (inverted diabolo). (e) Calculated out-of-plane magnetic field.}
  \end{center}
\end{figure}
}

\begin{figure}[h!]
  \begin{center}
    \includegraphics[clip,width=\linewidth]{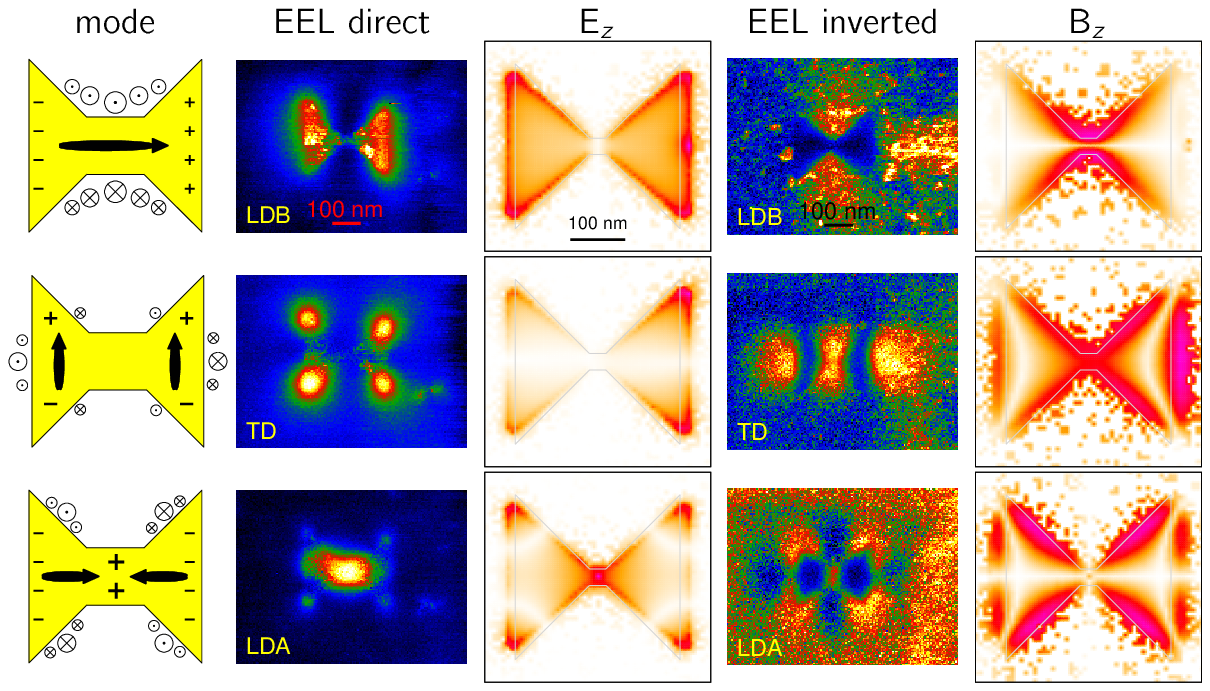}
    \caption{\label{figNO6} Experimental EEL maps and calculated fields for the LDB (1$^\mathrm{st}$ row), TD (2$^\mathrm{nd}$ row), and LDA (3$^\mathrm{rd}$) modes. (1$^\mathrm{st}$ column) Schematic representation of the mode in the diabolo antenna. The black arrow depicts the current with the thicker middle part representing the current antinode and the thinner border parts representing the current node. Magnetic field in the plane of the antenna is schematically represented by $\odot$ (outward field) and $\otimes$ (inward field) symbols with the size corresponding to the magnitude of the field. The node of current corresponds to the accumulation of the charge represented by $+$ and $-$ symbols. (2$^\mathrm{nd}$ column) Experimental loss intensity distribution. (3$^\mathrm{rd}$ column) Calculated out-of-plane electric field. (4$^\mathrm{th}$ column) Experimental loss intensity distribution for Babinet-complementary antenna (inverted diabolo). (5$^\mathrm{th}$ column) Calculated out-of-plane magnetic field.}
  \end{center}
\end{figure}

Finally, we address the question how to interpret the maps of the loss intensity and what to learn about the modes. We will first consider the case of LDB mode in the diabolo antenna. Figure~\ref{figNO6}(a) shows schematically how current oscillates and where the charge accumulates. Loss intensity recorded by EELS is proportional to the electric field projected to the trajectory of the electron~\cite{RevModPhys.82.209}, i.e., out of plane of the antenna. From the Gauss's law it follows that the strongest out-of-plane field occurs just near the areas of accumulated charge. Figure~\ref{figNO6}(b,c) indeed shows maxima of experimental loss intensity and calculated out-of-plane electric field at those areas. The interpretation of the loss intensity recorded for the inverted diabolo antenna (aperture) [Fig.~\ref{figNO6}(d)] is less straightforward as there is no simple current distribution of the related plasmonic mode. Instead, we will relate the loss intensity distribution of inverted antenna to the current distribution of the {\it direct} antenna. Loss intensity in the inverted antenna is proportional to the out-of-plane electric field, which is according to Babinet's principle proportional to the out-of-plane component of the magnetic field in the direct antenna. Indeed, magnetic field in the direct antenna according to the Amp{\`{e}}re's law circulates around the electric current with the maxima around the antinodes of the current [Fig.~\ref{figNO6}(a)]. Both the experimental loss intensity in the inverted antenna [Fig.~\ref{figNO6}(d)] and the calculated out-of-plane magnetic field in the direct antenna [Fig.~\ref{figNO6}(e)] exhibit maxima in the aread around the current's antinodes.

\del{
\begin{figure}[h!]
  \begin{center}
    \includegraphics[clip,width=\linewidth]{figure7.eps}
    \caption{\label{figNO7} (a) Schematic representation of the TD mode in the diabolo antenna. (b) Calculated out-of-plane electric field. (c) Calculated out-of-plane magnetic field.}
  \end{center}
\end{figure}

\begin{figure}[h!]
  \begin{center}
    \includegraphics[clip,width=\linewidth]{figure8.eps}
    \caption{\label{figNO8} (a) Schematic representation of the LDA mode in the diabolo antenna. (b) Calculated out-of-plane electric field. (c) Calculated out-of-plane magnetic field.}
  \end{center}
\end{figure}
}

Similar considerations can be repeated for TD mode and LDA mode or for bowtie antenna. Schemes of the charge oscillations together with the calculated electric and magnetic field distribution are shown in \del{Fig.~\ref{figNO7} for the TD mode and in Fig.~\ref{figNO8} for the LDA mode} the second and third line of Fig.~\ref{figNO6} for the TD and LDA mode, respectively. \del{Corresponding experimental EEL maps are shown in Fig.~\ref{figNO5}.} As in the case of LDB mode, bright spots in the loss intensity of direct antennas correspond to the strong out-of-plane component of the electric field related to the accumulated charge, and bright spots in the loss intensity of the Babinet-complementary antennas (apertures) correspond to the strong out-of-plane component of the magnetic field that circulates around antinodes of the current density.

\section*{Conclusion}

We have demonstrated a unique approach of independent engineering of individual LSP modes in composite plasmonic antennas based on varying the coupling between the components from capacitive to conductive. Taking bowtie and diabolo antennas as an example, we have shown that the longitudinal dipolar mode can be significantly modified (including switching between electric and magnetic hot spot) without affecting the higher-order modes. In combination with Babinet's principle allowing to engineer independently spectral and spatial properties of LSP modes, this represents a powerful tool for tailoring the properties of LSP modes for the phenomena and applications including Fano resonances, directional scattering, or enhanced luminescence.

\del{Further, we have shown the EELS characterization of direct antennas and their Babinet-complementary counterparts allows to retrieve both electric and magnetic near fields related to LSP modes, as well as the distribution of related charge and current oscillations.}

\section*{Methods}

\subsection*{Fabrication of antennas}

Plasmonic antennas for EELS measurement were fabricated by focused ion beam (FIB) lithography. First, we deposited 30~nm-thick gold layer on 30~nm-thick standard TEM silicon nitride membrane with the window size of $250\times250$~$\mu$m$^2$ (Agar Scientific). Gold layer was deposited without any adhesion layer by ion beam assisted deposition. Second, we performed FIB lithography in dual beam FIB/SEM microscope FEI Helios using gallium ions with the energy of 30~keV and ion beam current of 2.6~pA. We note that the energy (the highest available) and the current (the lowest available) are optimized for the best spatial resolution of the milling. Direct antennas were situated in the middle of a metal-free square (between $2\times2$~$\mu$m$^2$ and $4\times4$~$\mu$m$^2$), which is perfectly sufficient to prevent their interaction with the surrounding metallic frame~\cite{Krapek:15}. Inverted antennas were fabricated as isolated antennas with the distance between two nearby structures at least of 5~$\mu$m to prevent their collective interaction.

Plasmonic antennas for optical measurement were fabricated by electron beam lithography (EBL).

\subsection*{Electron energy loss spectroscopy}

EELS measurements were performed with TEM FEI Titan equipped with GIF Quantum spectrometer operated in monochromated scanning regime at 300~kV. Beam current was set to 0.4~nA and the FWHM of the zero-loss peak was around 0.1~eV. We set convergence angle to 10~mrad, collection angle to~10.4 mrad, and dispersion of the spectrometer to 0.01~eV/pixel. We recorded EEL spectrum images with the pixel size of 5~nm, while the number of pixels depended on the antenna size. Every pixel consists of 30 cross-correlated EEL spectra with the acquisition time around 25~$\mu$s for each spectrum. EEL spectra were integrated over several pixels around the positions of interest, background and zero-loss peak subtracted, and divided by the integral intensity of the whole spectrum to transform measured counts to a quantity proportional to the loss probability. EEL maps were obtained as energy-integrated intensity at the plasmon peak energy with the energy window of 0.1~eV divided pixel-by-pixel by the integral intensity of the zero-loss peak (integration window from $-0.5$~eV to 0.5~eV).

The relation of the measured loss probability to the near field of LSP modes is as follows. The probing electron transfers a part of its energy $\hbar\omega$ to a LSP mode in the specimen with the probability density (so-called loss probability) $\Gamma(\omega)$ reading~\cite{RevModPhys.82.209}
$$
\Gamma(\omega)=\frac{e}{\pi\hbar\omega}\int dt \mathrm{Re}
\left \{ \exp(-i\omega t) \bm{v} \cdot \bm{E}^\mathrm{ind}[\bm{r}_e(t),\omega] \right \}
$$
where $\bm{E}^\mathrm{ind}[\bm{r}_e(t),\omega]$ is the field induced by the electron moving with the velocity $\bm{v}$ at the position of the electron $\bm{r}_e(t)$. For the electron with the trajectory perpendicular to the sample (along the axis $z$) it is out-of-plane component of the field ($E_z$) that is relevant for the interaction. Correspondence between the loss probability and the electromagnetic local density of states projected along the electron beam trajectory has been thoroughly discussed~\cite{PhysRevLett.100.106804,C3CS60478K,PhysRevLett.103.106801}.

\subsection*{Optical measurements}

The far-field extinction ($1-T/T_\mathrm{ref}$) spectra were acquired using an infrared microscope (Bruker Hyperion 3000, 36$\times$ objective, $\mathrm{NA} = 0.5$) coupled to a Fourier-transform infrared spectrometer (Bruker Vertex 80V). Each spectrum was referenced to a bare substrate in the vicinity of the respective antenna array. 

\subsection*{Simulations}

In all simulations, the bow-tie and diabolo antennas have been represented by two gold triangles or triangular apertures (as shown in Fig.~\ref{figNO1}) of the height of 30~nm on top of 30-nm-thick silicon nitride membrane. The adhesion layer has been neglected. The dielectric function of gold was taken from Ref.~\cite{PhysRevB.6.4370} and the dielectric constant of the silicon nitride membrane was set equal to 4, which is a standard approximation in the considered spectral region~\cite{doi:10.1021/nl502027r}.

Electron energy loss spectra has been calculated with finite element method (FEM) using a commercial software COMSOL Multiphysics. 

\section*{Data availability}
The datasets analysed during the current study are available from the corresponding author on reasonable request.

\bibliography{ref,refHotspots,hotspot}

\section*{Acknowledgement} 
We acknowledge the support by the Czech Science Foundation
(grant No.~17-25799S), Ministry of Education, Youth and Sports of the Czech Republic
(projects CEITEC 2020, No.~LQ1601, and CEITEC Nano RI, No.~LM2015041), and
Brno University of Technology (grant No.~FSI/STI-J-18-5225).

\section*{Author information}
\subsection*{Contributions} 
V.K. conceived and coordinated research with help of T.S. Mi.H. and J.B fabricated the antennas, Mi.H. performed their electron spectrocopy with the assistence of M.S., F.L. performed optical spectroscopy, Ma.H. and A.K. performed numerical simulations. All authors were involved in the data processing and interpretation. V.K. and Mi.H. wrote the manuscript.  

\subsection*{Competing interests}
The authors declare no competing interests.

\end{document}